\documentclass{article}    

\usepackage{graphicx}

\title{\textbf{Consciousness: The rules of engagement}}  
\author{Richard Mould\footnote{Department of Physics and Astronomy, State University of New York, Stony Brook,
\mbox{New York} 11794-3800; http://nuclear.physics.sunysb.edu/ \~{}mould}}  
\date{}    
   
\begin{document}             

\maketitle              

\begin{abstract}

      	We examine the role of a conscious observer in a typical quantum mechanical measurement.    Four rules
are given that govern stochastic choice and state reduction in several cases of continuous and intermittent 
observation.  It is found that consciousness always accompanies a state reduction leading to observation, but its
presence is not sufficient to `cause' a reduction.  The distinction is clarified and codified by the rules that
are  given below.  This is the first of several papers that lead to an experimental test of the rules, and of the
``parallel principle" that is described elsewhere.     

\end{abstract}

\section*{Introduction}

A free particle interacts with a detector in such a way that it might be captured, or it might pass over the
detector without being captured.  During the interaction the quantum mechanical state of the system is given by
\begin{equation}
\Phi(t) = \psi(t)D_0 +D_1(t)
\end{equation}
where $\psi(t)$ is the incoming/scattered particle wave as a function of time.  This function is correlated with a
detector state $D_0$ that has not captured the particle.  The second component $D_1(t)$ is the detector state that
includes the captured particle.  $D_1(t)$ is initially equal to zero and increases in time, whereas $\psi(t)D_0$
is initially normalized and decreases in time.

 The first component in eq.\ 1 is really a time dependent expansion that describes a gradual transfer of momentum
from the scattered part of the wave to the detector, including y-components.  But because of the macroscopic
nature of the detector, these details can be ignored; so the detector state $D_0$ is approximated by a single
constant term that is factored out of its entanglement with the scattered wave.  With $D_0$ normalized, we have

\begin{displaymath}
\int\{\psi(t)^*\psi(t) + D_1(t)^*D_1(t)\} = 1
\end{displaymath}

	It may be objected that eq.\ 1 cannot be a superposition of quantum mechanical states because the detector states
are macroscopic.   Strictly speaking, each component in eq.\ 1 should have an environmental term attached, because
the two detector states affect the radiation or sonic field differently.  So the first component should read
$D_0E_0$, and the second component should read $D_1E_1$, where $E_0$ and $E_1$ are orthonormal environmental
states.  

\begin{displaymath}
\Phi(t) = \psi(t)D_0E_0 + D_1(t)E_1
\end{displaymath}
The cross terms are therefore zero when the environment is integrated out, reflecting environmental decoherence. 
This means that the detector states cannot locally interfere with one another, and this is often considered to be
the defining property of a macroscopic system.  It is what generally justifies calling a macroscopic object 
``classical" \cite{JZ,Gea}.  But non-interference is not really a universal property of macroscopic things, for
certain macroscopic systems at low temperatures have been shown to display interference effects as a result of
cryostatic isolation from the environment  \cite{JF, TL, TK, CW}.

The defining property of a macroscopic system is only that it is `big'.  It is no less quantum mechanical if it
lacks interference terms.  Since a possible phase relationship between the two components in eq.\ 1 is of no
importance to the problem being considered, that possibility is hereafter ignored together with the above
environmental terms.  In every other respect the detector is a quantum mechanical object that responds to a
quantum mechanical interaction with the particle.  I therefore continue to call the sum of components in eq.\ 1 a
quantum mechanical superposition\footnote{Joss and Zeh call the local system an ``improper mixture"(ref. 1).  I
refer to the total superposition because it is the system's quantum mechanical properties that I want to
explore.}.

\section*{The Conscious Observer}

My primary assumption is that conscious brains are legitimate components of a quantum mechanical system, and that
it is proper to ask how consciousness interacts with the system if it is present \emph{during} an ongoing
interaction.  Brains, conscious or not, are subject to the same principles of environmental decoherence as any
other macroscopic object, so they are similarly responsive to the methodology of quantum mechanics.  They just
don't display interference effects.  This treatment shares Everett's many-world assumption that conscious brains
are ordinary  quantum mechanical objects \cite{JCE}, but it differs in that the rules adopted below
allow only one of the many-world branches to be conscious at a time.

\begin{figure}[h]
\centering
\includegraphics[scale=0.8]{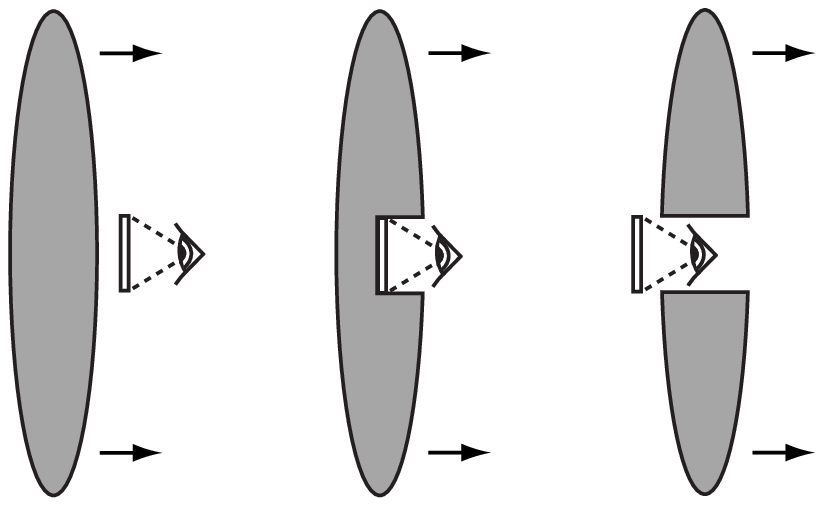}
\center{Figure 1}
\end{figure}

We begin by examining the above interaction when viewed by an observer as shown in fig.\ 1.  The first stage of the
figure shows the particle approaching the detector, where the particle is represented by the shaded area moving to
the right, and the detector is represented by the stationary white rectangle.  The observer is represented in the
figure by an eye that is viewing that surface.  

	The observer is engaged by radiation coming from the detector.  Equation 1 is therefore amended to include
normalized brain states of the observer $B_0$ and $B_1$ that are correlated with 
the detector states $D_0$ and $D_1$.\footnote{$B_0$ is entangled with $D_0$
and $B_1$ is entangled with $D_1$; so strictly speaking, they cannot be represented as the simple products
$D_0B_0$ and $D_1B_1$.  As before, this is an approximation that is justified by the macroscopic nature of these
states.}

\begin{displaymath}
\Phi(t) = \psi(t)D_0B_0 + D_1(t)B_1
\end{displaymath}
Since the time of Schr\"{o}dinger's cat it has been considered unphysical to imagine a `conscious' brain state in
superposition with something else.  But if the observer is consciously looking at the detector from the beginning,
and if the interaction proceeds for a finite time before the detector (possibly) captures the particle, then
during that time $B_0$ will be conscious \emph{and} will be continuously interacting with the detector.  So prior
to the particle being captured, $B_0$ will be conscious and $B_1$ will not be conscious.  I denote this difference
by underlining $B_0$ but not $B_1$, thereby amending the above equations to read 

\begin{equation}
\Phi(t) = \psi(t)D_0\underline{B}_0 + D_1(t)B_1
\end{equation}

I call $B_1$ a \emph{ready} brain state, which means that it is physiologically capable of becoming a conscious
state if and when the detector is stochastically chosen (i.e., if and when there is a capture).  There are
therefore three categories of brain states (i) conscious, (ii) unconscious, and (iii) ready.  The latter does not
have a classical meaning, for it occurs only in a quantum mechanical superposition.  The distinction between a
conscious brain state and a ready brain state is unavoidable if one accepts the possibility that brain states can
exist in quantum mechanical superposition, and that only one component can be conscious at a time.  The other
components would then be `ready'.

A ready state is not $un$conscious.  An unconscious state is physiologically incapable of consciousness as when the
observer is asleep or dead; whereas, a ready brain state is fully capable of consciousness when it is
stochastically ``realized".  Measuring any quantum mechanical variable produces an eigenvalue that is consciously
realized, as opposed to others that are not.  In this case, $B_1$ will not become ``real" unless and until it is
stochastically chosen to succeed $\underline{B}_0$.  That happens when the detector captures the particle.  

A brain state $B$ or $\underline{B}$ will be understood to be limited to the part of the cortex that can be set
into direct correspondence with conscious experience.  This excludes all those parts of the brain that are
involved in image processing.  These ``lower" parts will now be included in the detector, so the detector in
\mbox{eq. 2} is different than the one in eq.\ 1.

According to the von Neumann theory of state reduction, it is consciousness that initiates the reduction of a
quantum mechanical state \cite{vN}.  We see that that is not true, for the superposition in eq.\ 2 is not reduced
by the presence of a conscious brain state in one of its components.  However, consciousness is associated with the
state reduction that occurs when the detector records a particle capture.

\section*{Possible Outcomes}

Suppose the particle is detected and observed before the end of the interaction.  This will require a stochastic
choice based on quantum mechanical probabilities, and we will say that that happens at a time $t_{sc}$.  The state
in eq.\ 2 will then be reduced, giving 

\begin{equation}
\Phi(t \ge t_{sc}) = D_1\underline{B}_1
\end{equation}
 In this case, the superposition in eq.\ 2 is reduced so that only its second component survives, consciousness is
transferred to the brain state
$\underline{B}_1$, and the interaction is terminated.

	The other possible outcome is that there is no particle capture during the time of the interaction.  In that
case eq.\ 2 will remain undisturbed, even after the interaction is complete at time $t_f$.  The system will
therefore remain in a \emph{residual superposition} given by

\begin{equation}
\Phi(t>t_f) = \psi(t)D_0\underline{B}_0 + D_1(t_f)B_1
\end{equation}
Since the observer is only conscious of the first component in this equation, the existence of the second (not
conscious) component is of no empirical concern, even when there is no longer any possibility of a capture.  Of
course, the form of eq.\ 4 is potentially important to another observer who, after $t_f$, observes the system over
the shoulder of the primary observer.  We will see in another section that the second component in eq.\ 4 will be
eliminated by the mere `interactive' presence of a second observer.  In addition, we will see it 
eliminated by the primary observer himself when his conscious attention drifts in any way from the state
$\underline{B}_0$.  However, for the moment we will leave eq.\ 4 as it stands.  It is not empirically wrong.

\section*{The Rules}

I accept that brain states can exist in superposition with themselves, but I do not accept an Everett-like
plurality of simultaneous conscious components.  It is therefore necessary to find the general rules that govern
the relationship between conscious and ready brain states, consistent with the requirement that only one conscious
state of an observer can be realized at a time.  In particular, the rules must tell us how and when a reduction
occurs so as to bring about the outcomes in the previous section.  I propose four rules that I believe are
sufficient to describe all the interactions that are considered in this and in subsequent papers.  These rules are
\emph{ad hoc} in that they are made to fit the cases studied.  It is my belief that if and when we find a proper
theory of the relationship between consciousness and matter, these rules, or rules like them, will naturally
emerge.    

	The first three rules are adequate to deal with the outcomes in eqs.\ 3 and 4, as well as several other situations
that are described in the following sections.  I therefore begin by introducing just these three.  The fourth rule
will be offered when it becomes clear that the first three by themselves are unable to resolve a particular
difficulty that arises.  

A central concept in rule (1) is \emph{probability current} $J$, which is defined to be the time rate of change of
square modulus.  Probability in this treatment is introduced only through the current $J$.    

\vspace{0.4 cm}

\noindent
\textbf{Rule(1)}: \emph{For any subsystem of n components in an isolated system with a square modulus equal to s,
the probability per unit time of a stochastic choice of one of those components at time t is given by
$(\Sigma_nJ_n)/s$, where the net probability current $J_n$ going into the $n^{th}$ component at that time is
positive.  }

\vspace{0.4 cm}

Rule (1) puts no limit on the kind of state that is chosen, so long as $J > 0$.  It will apply to any
interaction in any representation.  There is no need to say that it applies only to irreversible interactions, or
to components representing macroscopic objects, or to objects that lack the possibility of interference.   

I cannot say why Nature should harbor such a triggering device any more than I can say why Nature should make use
of intrinsic probability.  But if probability is natural to a physical system, then so is stochastic choice.  With
this understanding, I would say that rule (1) adds nothing that is not already implicit in standard quantum
mechanics; except that the underlying notion of probability is shifted from the square modulus to the probability
current. In rule (2), an \emph{active} brain state is one that is either conscious or ready.  It is actively engaged
with the apparatus.

\vspace{0.4 cm}

\noindent
\textbf{Rule(2)}: \emph{If the Hamiltonian gives rise to new components that are not classically continuous with
the old components or with each other, then all active brain states that are included in the new components will be
ready brain states.}  

\vspace{0.4 cm}

This rule provides for the introduction of ready brain states.  For example, if a conscious brain state
$\underline{B}$ interacts with an apparatus state $A_k$, then the  product of independent states
$A_k\underline{B}$ will lose amplitude to an entangled state $A_kB_k$. That is,   
\begin{displaymath}
A_k\underline{B} \hspace{.3cm}\mbox{becomes}\hspace{.3cm}A_k\underline{B} + A_kB_k 
\end{displaymath}
As before, I write the result $A_kB_k$ as a simple product for ease of recognition.  It is a good approximation for
these macroscopic states.  The emerging state $B_k$ must be a ready brain
state according to rule (2). 

In another example, an unknown brain state $X$ interacting with a superposition of
apparatus states $A_r + A_s$, will give rise to another superposition  
\begin{displaymath}
(A_r + A_s)X \hspace{.3cm}\mbox{becomes}\hspace{.3cm}(A_r + A_s)X + A_rB_r + A_sB_s 
\end{displaymath}
where the emerging states $B_r$ and $B_s$ must be ready brain states.  In the present paper, all different brain
states will be discrete.  In a future paper we will consider  brain states to be a continuum
in which close neighborhood states are not discrete.        

The third rule describes the collapse of the wave function associated with measurement. This wave collapse, or
state reduction, does \emph{not}  interrupt the Schr\"{o}dinger process, and it does not
modify the Hamiltonian or withhold its continuous application.  I regard the collapse as an
abrupt change in the \emph{boundary conditions} on the solutions of the Schr\"{o}dinger equation, and that is
all.  This is what an observation does.  It adds new information.  That information abruptly changes the physical
context, and hence, it abruptly changes the Schr\"{o}dinger solution.  It does not change the evolutionary
process itself.  

\vspace{0.4 cm}

\noindent
\textbf{Rule(3)}: \emph{If a component AB is stochastically chosen, where A is the total amplitude of a total ready
brain state B that emerges from an interaction, then B will become conscious and all the other components will be
immediately reduced to zero.}

\vspace{0.4 cm}

This rule does not provide for renormalization after a stochastic choice.  One might include that
requirement for convenience, but it is not necessary.  We will continue to normalize $B_0$, $B_1$, and $D_0$, but
not $\Phi(t)$ following a stochastic collapse.  If rule (1) is followed as stated, the probabilities will work out
correctly without a normalization requirement.  

 Rule (3) does \emph{not} say that a ready brain state must be present for there to be a state reduction.  One
might suppose that a state reduction can occur in \mbox{eq.\ (1)} without the presence of a conscious observation
of any kind, but I do not believe that will ever happen.  I believe that a stochastic trigger will have no
consequence unless, through rule (3), a ready brain state is both available and stochastically chosen, followed by
a reduction that favors the chosen state.  My view is similar to one expressed by von Neumann and Wigner
that attaches fundamental importance to conscious observation.  

All of the examples that I present in these papers reflect this understanding because they all deal with reduction
in the presence of an observer.  If the reader believes that this thinking is to narrow, he is welcome to
generalize \mbox{rule (3)} to describe a reduction in the absence of an observer.  There are many problems
associated with a generality of this kind, principally the question of representation and basic eigenstates.  I
dispose of these questions by specializing the rule as I do.  In any case, there can be nothing wrong with my
limiting this study to cases in which an observer is present.  We know what the conscious experience is in
these cases, so we are in possession of the experiential `facts'.  A broader interpretation of rule (3) should not
invalidate these facts or contradict the way I put them into a formal framework.   

The `possible outcomes' of the previous section follow from these rules.  \mbox{Rule (1)} explains why the first
(conscious) component in eq.\ 2 is not selected for a reduction that eliminates the second (non-conscious)
component.  It is because current flows out of the first component making $J < 0$.  Rule (2) insures that
\mbox{eq.\ 2} is not a superposition of two conscious states.  Rule (3) tells us why the conscious state
$\underline{B}_1$ survives the collapse resulting in eq.\ 3.  It is because the positive flow of current into the
ready state $B_1$ permits a stochastic hit, and that causes the state to become conscious and to be the sole
survivor of the reduction.    These rules therefore describe the expected outcomes of the interaction.  They
explain much more as will be shown.

I make no attempt to physiologically define a conscious brain state or a ready brain state; but however that is 
done, their role as basis states is a matter of importance to quantum ``epistemology", even in the Copenhagen
interpretation.  Stapp makes the point that according to Copenhagen, the observer ``who stands outside of the
Hilbert space structure decides how he will set up the experiments, and this decision fixes the `basis'
(states)"\cite{HS}.  Something beyond the Schr\"{o}dinger process is surely necessary to fix the basis, and that
is taken by Copenhagen to be the observer's experience of the apparatus.  In the present treatment the basis is
fixed in the same way, and is made explicit in rule (3).  Only through \mbox{rule (3)} does the trigger have
subjective reality implications, and that happens only when the trigger hits a ready brain state that is created
by rule (2).  When a choice like that is made, the chosen state becomes a conscious state, establishing a new
boundary condition on Schr\"{o}dinger's equation.  

It is apparent that consciousness is an important player in the above interaction.  However, the rules
do not  represent it as having a `causal' influence of any kind.  The stochastic choice of a ready brain state is
said to do two things.  It makes that state conscious \emph{and} reduces all others to zero.  Consciousness is said to appear in
these circumstances, but it will have no effect of its own on the physical system. The rules therefore preserve
the traditional epiphenomenal nature of consciousness.  Of course,  a \mbox{rule (3)} reduction  depends on the
existence of a ready brain state.  To that extent, the observer is an essential part of the methodology; and that
is a departure from  standard interpretations in the direction suggested by von Neumann and Wigner.  

Future papers will add two more rules to the four given in this paper.  The additional rules are specific to a
model of the brain that takes account of the continuous nature of brain states.  The first of these
addditional rules merely expands \mbox{rule (3)} to accommodate the continuum case.  The final rule that we add 
will give consciousness a causal influence that should be measurable in principle.  Therefore, the structure that
is developed here has eventual consequences that are experimentally testable in principle.  See the final section
of this paper.

\section*{A Terminal Observer}

	Going back to the bare interaction in eq.\ 1, the question is: do the above rules adequately describe the expected
results of a terminal observation in this case?  Are they adequate to the usual findings in a physics laboratory
where an observer only comes into the picture when the interaction is complete?  After the final pass-over time
$t_f$, the system in eq.\ 1 stabilizes  at

\begin{displaymath}
\Phi(t>t_f) = \psi(t)D_0 + D_1(t_f)
\end{displaymath}
This superposition will be reduced when it interacts with the observer after time $t_f$.  As before, a ready state
$B_0$ will become correlated with the `no capture' detector state $D_0$, and $B_1$ will become correlated with the
`capture' detector state $D_1$.  The state $X$ given below in eq.\ 5 is an unknown brain state of the observer
prior to his turning his attention to the detector.  State $X$ will not be a ready brain state\footnote{If the
terminal observer's initial $X$ state is either conscious or unconscious, it is that state that will be
``converted" to the new experience when the interaction begins, and not an associated ready brain state.  Rule (4)
will provide a reason why $X$  \emph{cannot} be a ready brain state.}.  According to rule (2), the system after
the initial  interaction at time $t_{ob} > t_f$ is  
\begin{eqnarray}
\Phi(t\ge t_{ob}>t_f) & = & \psi(t)D_0X + D_1(t_f)X \\
                  & + & \psi'(t)D_0B_0 + D_1'(t_f)B_1  \nonumber
\end{eqnarray}
where conscious states do not yet appeared, and the primed states in the second row are equal to zero at the moment
of observation.  Immediately after that, the probability currents of the unprimed components will be negative, and
the probability currents of the primed components will be positive

\begin{displaymath}
J_0 < 0 \hspace{0.2cm} \mbox{and}\hspace{0.2cm} J_1 < 0  \hspace{1.5cm}      J_0' > 0\hspace{0.2cm} \mbox{and}
\hspace{0.2cm}J_1' > 0
\end{displaymath}

Inasmuch as the initial state is normalized, rule (1) with $n = 1$ tells us that the probability of a stochastic
hit in time $dt$ is equal to $J_0'dt$ for the first primed component in eq.\ 5, and $J_1'dt$ for the second. 
According to rule (3), the first hit that occurs will cause the affected brain state to become conscious, yielding
either $\underline{B}_0$ or $\underline{B}_1$, and will reduce the other states to zero.

We also know that there will be at least one stochastic hit on the system.  Rule (1) with $n = 2$ gives  

\begin{displaymath}
\int[J_0' + J_1']dt =  1
\end{displaymath}
where currents $J_0'$  and  $J_1'$ are assumed to flow until the observerÕs physiological interaction is
complete. 

The integrated probability of 1.0 insures at least one stochastic hit.  Furthermore, there will be no more than
one hit because the reduction resulting from the first hit will reduce all other components to zero.  A second hit
on the component that is already conscious is meaningless.  Therefore, dropping the prime and the time dependence,
the final state of the system after reduction at $t_{sc}$ will be either
\begin{eqnarray}
&D_1\underline{B}_1\\
\mbox{or}&\psi(t \ge t_{sc} > t_{ob} > t_{f})D_0\underline{B}_0\nonumber
\end{eqnarray}
depending on the stochastic choice.  Both reductions occur with a total probability given by their
square moduli at time $t_{f}$.

	The terminal result in eq.\ 6 is the one that is most familiar in a physics laboratory.  The first row is
identical with eq.\ 3, and the second is identical with eq.\ 4 so far as the observer is concerned, inasmuch as he
is unaware in eq.\ 4 that he is part of a residual superposition.  These results are therefore in (empirical)
agreement with the possible outcomes described above for an observer who watches the interaction from the
beginning.

\section*{An Intermediate Observer}

Suppose the observer first looks at the detector after the interaction has begun, but before the particle has
stopped interacting with the detector.  The observation will then occur sometime after $D_1(t)$ in eq.\ 1 has
acquired a finite value, and while current into that component is still positive.  The equation is then 
\begin{eqnarray}
\Phi(t_f > t\ge t_{ob}) &=& \psi(t)D_0X + D_1(t)X \\
&+& \psi'(t)D_0B_0 + D_1'(t)B_1 \nonumber
\end{eqnarray}
where $B_0$ and $B_1$ are the ready brain states of the observer, and where $\psi'(t)$ and $D_1'(t)$ are zero at
$t_{ob}$.  State $X$ is again the unknown brain state of the observer prior to turning attention to the detector;
where again, it cannot be a ready state (footnote 4).

Current flowing vertically from the first row to the corresponding component in the second row is the result of
the physiological interaction.    It might very well give rise to a stochastic choice of $D_1'B_1$ at some time
$t_{sc1}$, in which case the entire process will be brought quickly to an end.  Dropping the prime and the time
dependence, only the component $D_1\underline{B}_1$ would then survive. 
\begin{equation}
\Phi(t\ge t_{sc1}>t_{ob}) = D_1\underline{B}_1
\end{equation}
This reflects the possibility that the particle has already been captured when the observer makes his appearance.

The other primed component might be stochastically chosen, and that would select $\psi(t)D_0\underline{B}_0$
(dropping the prime) at a time $t_{sc0}$ in the middle of the primary interaction.  A mid-stream collapse of this
kind is not unphysical.  It amounts to starting again at a time when the function $\psi(t)$ has  passed over the
detector to some extent without a particle capture.  A physicist might very well look at the detector in the
middle of an interaction, note that the particle has not yet been detected, and determine the initial conditions
to be given by the solitary component $\psi(t)D_0\underline{B}_0$ starting at the new time
$t_{sc0}$.  The interaction would proceed from there to give
\begin{displaymath}
\Phi(t_f >t \ge t_{sc0}>t_{ob}) = \psi(t)D_0\underline{B}_0 + D_1''(t)B_1
\end{displaymath}
where $D_1''(t)B_1$ is equal to zero at $t_{sc0}$.  If there is another stochastic hit at a time $t_{sc1'}$, the
system will become
\begin{equation}
\Phi(t\ge t_{sc1'}>t_{sc0}>t_{ob}) =  D_1\underline{B}_1
\end{equation}

Equations 8 and 9 are two separate routes to a particle capture.  The total probability of their occurrence is
equal to the probability that there was a capture prior to the intermediate observation, plus the probability of a
subsequent capture.  

The remaining possibility is that the interaction terminates without a second stochastic hit at $t_{sc1'}$. 
Dropping the double prime, the residual superposition
\begin{equation}
\Phi(t>t_f > t_{ob}) = \psi(t)D_0\underline{B}_0 + D_1(t_f)B_1
\end{equation}
will then persist beyond $t_f$, and will last indefinitely as it did in eq.\ 4.

	There are therefore two consequences of an intermediate observation.  The observer will either experience a
capture (eqs.\ 8 and 9), or he will experience no capture and will continue until the end of the interaction as
part of the residual superposition in eq.\ 10.  The observer's experience after $t_{ob}$ is therefore identical
with that of an observer who was on board from the beginning.  As in that case, we claim and will show below that
the second component in eq.\ 10 will be eliminated by a second observer, or by drift consciousness of the primary
observer.

\section*{An Outside Terminal Observer - Rule (4)}

	If another observer looks at the apparatus at some time $t_{ob}$ after the interaction has terminated in the
residual superposition of eq.\ 4 or eq.\ 10, we will have according to rule (2)
\begin{eqnarray}
\Phi(t \ge t_{ob}>t_f) &=& \psi(t)D_0\underline{B}_0X + D_1(t_f)B_1X\\
&+& \psi'(t)D_0\underline{B}_0B_0 + D_1'(t_f)B_1B_1\nonumber 
\end{eqnarray}
where the primed components are equal to zero at $t_{ob}$.  The brain states of two observers will appear as a
product such as
$B_1X$, where the first refers to the first observer and the second refers to the second observer.  In this case,
$B_1$ is the ready brain state of the primary observer, and $X$ is the unknown brain state of the outside observer
prior to his interacting with the detector.  The state $X$ could be either conscious or unconscious, but not ready
as per \mbox{footnote 4}.  Equation 11 parallels eq.\ 5, except that the outside observer is now in the process of
coming on board with the primary observer.

The second component in eq.\ 11 seems to contribute positive current to the fourth component, and that could result
in a stochastic hit on the capture state $D_1'(t_f)B_1B_1$ \emph{after} the particle has passed over the detector. 
This is not possible, for the primary observer cannot witness a capture after the primary interaction has been
terminated.  The rules apparently allow an anomalous capture of this kind, and we must to do something to insure
that that does not happen.  \mbox{Rule (4)} prevents it from happening.

\vspace{0.4 cm}

\textbf{Rule(4)}\emph{A transition between two components is forbidden if each is an entanglement containing
 a ready brain state of the same observer.}

\vspace{0.4 cm}

An interaction between a ready brain state and a conscious or unconscious state is not affected by this rule. 
So the primary interaction between the first and the second component in eq.\ 11 is preserved, as is the
physiological interaction between the first and the third components.  However, rule (4) states that a primary
interaction is not possible between the third and the fourth components, and that a physiological interaction is
not possible between second and the fourth components.  In addition, there is no term in the Hamiltonian that
directly connects the first and the fourth components.  Therefore, the fourth component $D_1'(t_f)B_1B_1$ in eq.\
11 does not belong there at all, for it cannot interact with any of the other components.  

The fourth rule does not just say that there is no interaction between ready brain states.  That mild requirement
would allow a transition from $X$ to  $B_1$ ($2^{nd}$ to the $4^{th}$ component), or from 
$\underline{B}_0$ to  $B_1$ (3$^{rd}$ to the 4$^{th}$ component).  The rule is much stronger.  It \emph{forbids}
any transition that carries a ready brain state into itself, or into another ready brain state of the same
observer.

	Rule (4) solves the anomaly problem.  The absent component $D_1'(t_f)B_1B_1$ cannot possibly be chosen, so the
there will not be an anomalous capture of the particle after $t_f$.  It is also too late (after $t_f$) for a
transition to the second component.  Therefore, the third component is the only candidate for stochastic choice.  
Furthermore, rule (1) assures that it \emph{will} be chosen.  

The original  state in eq.\ 11 is $\Phi(t_0) = \psi(t)D_0\underline{B}_0X$, and current goes from that
state into the second and third components of eq. 11. From rule (1) with $n = 2$ we have
\begin{displaymath}
\int[J_x + J_0]dt =  1
\end{displaymath}
where currents $J_x$  and  $J_0$ flow into $D_1(t)B_1X$  and $\psi'(t)D_0\underline{B}_0B_0$.  The time integral
extends from $t_0$ to the end of the physiological interaction when the original state is entirely depleted.  Since
it is stipulated that the second component $D_1(t)B_1X$ in eq.\ 11 was not chosen during the primary interaction, 
the third component $\psi'(t)D_0\underline{B}_0B_0$ must be chosen at some time $t_{sc}$.    

From rule (3), the third component will then become  $\psi(t\ge t_{sc} > t_f)D_0\underline{B}_0\underline{B}_0$,
and  the interaction will be complete with the outside observer  on board with the primary observer.  This has the
effect of eliminating the superfluous (capture) component, so the system is no longer a residue superposition. 
And most important, there will be no anomalous capture. 

In a similar way, \emph{any} residual superposition can be reduced by the presence of an outside observer, thereby
removing the superfluous component in the superposition.  The rule also explains why the unknown state $X$
appearing in eqs.\ 5, 7, and 11 cannot be a ready brain state as claimed in footnote 4.  It is because a ready
brain state is forbidden by rule (4) to deliver current to the primed states appearing in the second row of each of
those equations.  In the second paper in this series, rule (4) forbids other anomalies that are different in kind
from the one described above.

There are now three features of a ready state that distinguish it from a conscious state.  (i) it is
not-conscious, (ii) it can be made conscious by a stochastic hit, and (iii) it cannot interact with others of its
kind.  It is not my general intention to propose physiological mechanisms, but there is one suggestion that may
help understand what might be going on.

Imagine that each ready brain state of an observer includes a physiological feature that I will call a \emph{stop},
and require that an interaction is forbidden between any two states that possess a stop (i.e.,
between two ready brain states of an observer).  Current can flow into or out of a stopped state, but it cannot
flow between two such states within one observer.  Imagine also that this same physiological difference between a
ready state and a conscious state is sufficient to explain the inability of a ready state to be conscious.  And
finally, imagine that when this state is stochastically chosen the stop is disabled, and this allows the state to
become conscious, and to interact more normally with other states.   Although I do not explicitly propose this
mechanism, the conjecture illustrates how a correct understanding of the rules governing psycho-physical
interactions might help to narrow the constraints on the Hamiltonian of a conscious brain.     

	Reviewing the effect of rule (4) on previous cases, we note that it produces no change in the terminal
observation of eq.\ 5, or in the intermediate observation of eq.\ 7, except to validate the claim that the state
$X$ on the first row of these equations cannot be a ready brain state.  We did not previously consider the
possibility of a stochastic choice arising from a current flow from the third to the fourth component of eq.\ 7. 
This possibility is now explicitly forbidden by \mbox{rule (4)}.

\section*{An Intermediate Outside Observer}

Suppose an outside observer makes an observation during the primary interaction, and before a particle capture,
looking over the shoulder of the primary observer in eq.\ 2.  If $t_{ob}$ refers to the time of this 
observation, then the moment the outside observer interacts with the detector, the system becomes 
\begin{eqnarray}
\Phi(t_f > t \ge t_{ob}) &=& \psi(t)D_0\underline{B}_0X + D_1(t)B_1X\\
&+& \psi'(t)D_0\underline{B}_0B_0 \nonumber 
\end{eqnarray}
where there is no possibility of a fourth component because of rule (4).  The primed component is again equal to
zero at $t_{ob}$.  

	If there is a stochastic hit on the second component in eq.\ 12 at time $t_{scx}$, the result will be 
\begin{displaymath}
\Phi(t_f >t = t_{scx}>t_{ob}) =  D_1(t)\underline{B}_1X
\end{displaymath}
This corresponds to a particle capture that occurs before the physiological interaction is complete.  Since that
interaction is still in engaged, this will lead directly to 
\begin{displaymath}
\Phi(t_f > t \ge t_{scx} > t_{ob}) =  D_1(t)\underline{B}_1X + D_1'(t)\underline{B}_1B_1
\end{displaymath}
where $D_1'(t)\underline{B}_1B_1$ is equal to zero at $t_{scx}$.

The only thing that can follow is another stochastic hit at $t_{sc1}$, giving
\begin{equation}
\Phi(t \ge t_{sc1} > t_{scx}) =  D_1\underline{B}_1\underline{B}_1
\end{equation}
thereby completing the measurement.      

If, on the other hand, the third component in eq.\ 12 is the first to be stochastically chosen at $t_{sc0}$, the
reduction will be
\begin{displaymath}
\Phi(t_f > t = t_{sc0} > t_{ob}) =  \psi(t)D_0\underline{B}_0\underline{B}_0 
\end{displaymath}
where the prime on $\psi(t)$ is dropped.  Since the primary interaction is still affective, this will become 
\begin{equation}
\Phi(t_f > t \ge t_{sc0} > t_{ob}) =  \psi(t)D_0\underline{B}_0\underline{B}_0 + D_1''(t)B_1B_1
\end{equation}

If there is a second stochastic hit before the particle interaction is complete at $t_f$, the result will be
identical with eq.\ 13.  However, if there is no capture before $t_f$, the residual superposition 
\begin{equation}
\Phi(t > t_f) =  \psi(t)D_0\underline{B}_0\underline{B}_0 + D_1''(t_f)B_1B_1
\end{equation}
will remain in place as it did in eqs.\ 4 and 10.  As in those cases, the second component serves
no further purpose at this time, and will be removed by the appearance of a third observer as was previously shown,
or by drift consciousness as will be shown below.

\section*{Drift Consciousness}

	Even though an observer may have a steady eye, his attention is bound to drift slightly to neighboring brain
states that continue to interact with the detector.  Suppose the observer in eq.\ 2 lets his mind wander to a
neighboring brain state that still allows him to be aware of $D_0$, but not in the same way.  The neighboring
state may regard the detector from a slightly different angle, or it may include qualitative differences that have
more to do with the observer's emotional state than anything else.  In any case, a new brain state will become
available through a physiological process that phases out $\underline{B}_0$ and gives rise to a new brain state
that also interacts with the detector.  The Hamiltonian will direct the drift that goes from the original brain
state to the neighboring states represented as  
\begin{eqnarray}
\Phi(t> t_0) &=& \psi(t)D_0\underline{B}_0 + D_1(t)B_1 \\
&+& \psi'(t)D_0B_{0a} +\psi''(t)D_0B_{0b} \nonumber\\
&+& \psi'''(t)D_0B_{0c}\nonumber + \mbox{etc.}
\end{eqnarray}
where $B_a$, $B_b$, $B_c$, etc. are qualitatively different from the given brain state.  

Probability current will flow out of the first component in eq.\ 16 into all of the other components.  Rule (4)
forbids current flow to captive states of the alternative brain states.  If one of the primed components is
stochastically chosen at some time
$t_{sc}$, there will be a reduction following rule (3).  Assuming that the double primed component is chosen, we
will have
\begin{equation}
\Phi(t_{sc}) = \psi(t)D_0\underline{B}_{0b}
\end{equation}
where the primes are now dropped.  The drift process will begin again starting at the new time $t_{sc}$, giving 
\begin{eqnarray*}
\Phi(t\ge t_{sc}) &=& \psi(t)D_0\underline{B}_{0b} + D_1'(t)B_{1b}\\
&+& \psi''''(t)D_0B_{0d} + \mbox{etc.}
\end{eqnarray*}
where all but the first component are zero at time $t_{sc}$.  So a system that begins with the first component
of eq.\ 16, becomes $\psi(t)D_0\underline{B}_{0b}$ in eq.\ 17 at some later time.  The effect of drifting is
therefore to renew the system at a new drift site at a new time, and to continue the `primary' interaction stating
at that time.  

	If eq.\ 16 were a residual superposition, the drift result would be eq.\ 17 without the possibility of a new
capture state such as $D_1'(t)B_{1b}$.  This means that drifting will eliminate the superfluous component of a
residual superposition as previously claimed.  We have seen that a second observer has that effect; and
apparently, the primary observer can do that all by himself.  

	When I speak of the above physiological drifting I do not mean the eye motion that produces a jitter of the
retina with respect to a visual image.  That motion is part of the image processing that is considered here to be
included in the detector, and is not perceptible in consciousness.  A brain state denoted by $\underline{B}$ or $B$
in this paper refers to a higher cortical state in which all image processing is complete, so the above result can
be set into direct correspondence with conscious awareness.  It is the variations in this direct awareness that I
say results in drift consciousness.  

On this drift model, the neighborhood states in \mbox{eq.\ 16} are finitely separated from one another. Although
the ready brain states in that equation are distinct, they evolve in an orderly quantum mechanical way that is
governed by the Hamiltonian; whereas, conscious states appear discontinuously when there is a stochastic switch
from one brain state to the next.  In a future paper we will consider the case in which neighboring brain states
are differentially close together.

\section*{Phantom Component}

It has now been shown that the second component of a residual superposition can be washed away by the presence of
another observer, or by drift consciousness on the part of the primary observer.  That component has rule (1)
physical significance as long as current flows into it from the first component; but the moment that current
stops, it is no longer a physically meaningful quantity.  It becomes a \emph{phantom} component of the
superposition.  It certainly does have a non-zero square modulus, but that is not interpreted as
`probability' in this treatment.  Of course, a physicist may want to integrate current flow to calculate total
probability, but the validity of that procedure does not mean that Nature gives intrinsic meaning the absolute
value of square modulus.  In the case of a phantom component, square modulus may be said to represent the
probability that the component \emph{might have been} stochastically chosen, but was not.

\section*{Drifting Away}

	If the observer drifts into unconsciousness by some process such as falling asleep, then the process will begin
in the same way described above.  The difference is that the Hamiltonian will lead the system into a part of the
brain that is not capable of supporting ready brain states, and cannot become conscious.  In that case, the
Hamiltonian will drive the ``last" conscious component to zero at a time $t_u$, insuring a decisive exit of
consciousness.  For instance, suppose that all of the primed brain states in eq.\ 16 are unconscious.  If the
Hamiltonian drives the first row to zero before the capture state $D_1(t)B_1$ can be stochastically chosen, then
the remaining system will consist of the lingering superposition
\begin{displaymath}
\Phi(t_u) = \{\psi'(t)D_0 + \psi''(t)D_0 + \psi'''(t)D_0 + \mbox{etc.}\}U
\end{displaymath}
where $U$ is the newly acquired unconscious state of the observer\footnote{Factoring out $U$ is an approximation,
inasmuch as the $U$ of each component is entangled with its own detector/particle state.}. 
This leaves the system in an elaborate superposition of detector states.  Since all the times in this equation are
equal, it can be simplified to read
\begin{displaymath}
\Phi(t \ge t_u) = \{\psi(t)D_0 + D_1'(t)\}U
\end{displaymath}
where $D_1'(t)$ is equal to zero at $t_u$.  This equation is identical with eq.\ 1, except that the unconscious
observer is now independent of the interaction.

\section*{Fast Interaction}

Nothing has been said about the primary interaction time compared to the physiological interaction time.  To take
account of a fast primary interaction, a more complete description of the detector is necessary.

As previously explained, the detector includes the observer, except for the part of the cortex that can be set
into direct correspondence with conscious experience.  It includes all image processing.  The detector therefore
requires at least physiological time to do its job.  Let it be represented by $D(\alpha, \beta, \gamma, . . .  ,
\omega)$ where $\alpha, \beta, \gamma, . . .  , \omega$ are internal variables.  In that case, current flowing into
the detector will first cause $\alpha$ to pulse, then $\beta$, and etc.  When this input is `impulsive' it will be
carried along by the other variables, spreading as it goes, and arriving at the ready brain
site long after the highly localized and fast moving particle has passed by the detector.  Therefore, it is not
unreasonable to suppose that all of the primary interaction times in the examples in this paper (governing the
horizontal flow) are executed in physiological time, or longer.

\section*{A Co-Observer}

Finally, consider what would happen if a co-observer participated in the interaction with the original observer. 
Let the scintillation area of the detector be divided into two parts.  We then have two separate brains,
where the first brain in each component of eq.\ 18 looks at the first detector area, and the second brain looks at
the second area.  As soon as the detector becomes entangled with the brain states of each observer, the equivalent
to eq.\ 2 is
\begin{equation}
\Phi(t) = \psi(t)D_{00}\underline{B}_0\underline{B}_0 + D_{10}B_1B_0 +
D_{01}B_0B_1
\end{equation}
where the detector state $D_{00}$ displays no captured particle, $D_{10}$ displays the captured particle in the
first area, and $D_{01}$ displays the captured particle in the second area.  The dual brain state
$\underline{B}_0\underline{B}_0$ represents the first observer experiencing no scintillations in the first area of
the detector, and the second observer experiencing no scintillations in the second area of the detector.  If
either of the dual ready brain components in eq.\ 18 is stochastically chosen at time $t_{sc}$, then both states
will qualify under rule (3) to achieve consciousness.  This will lead to either
\begin{displaymath}
\Phi(t>t_{sc}) = D_{10}\underline{B}_1\underline{B}_0 \hspace{.8cm}\mbox{or}\hspace{.8cm}\Phi(t>t_{sc}) =
D_{01}\underline{B}_0\underline{B}_1        
\end{displaymath}
If the superposition in eq.\ 18 survives the interaction, it will be reduced to the state
$\psi(t)D_{00}\underline{B}_0\underline{B}_0$ as a consequence of a third observer or drift consciousness on the
part of either one of the observers.  

	The above results can be generalized to apply when there is more than one particle in the wave $\psi(t)$.  In that
case, components with detector states of the form $D_{mn}$ will appear in product with the particle wave diminished
by $m + n$ particles, where $n$ is the number of particles striking the first scintillation area, and $m$ is the
number striking the second scintillation area.

\section*{Discussion}

	The purpose of this paper is to describe what must happen when a conscious observer witness a quantum mechanical
interaction, and to find the simplest rules that codify the results in different situations.  The underlying
assumption is that macroscopic brains (conscious or not) are legitimate components of a quantum mechanical
superposition, and that Everett's thesis is not a consideration.  To accomplish this, four rules are given that
describe how brain states affect the reduction of quantum mechanical systems.  

These rules do not explicitly require that only one brain state of an observer can be conscious at one time, but
they do manage to bring about that result.  The absurdity of indefinitely many-worlds of consciousness is thereby
avoided.  The many-world feature is here transferred to ready brain states that perform a non-trivial function in
this treatment.  These states are on standby, ready to take over the role of consciousness if and when they are
stochastically chosen.  When chosen, they contribute ``real" boundary conditions that govern the solutions
of Schr\"{o}dingerÕs equation.  

In a following paper I apply these four rules to the Schr\"{o}dinger cat experiment, and they give the normally
expected results \cite{RM1}.  This case also allows us to consider how the rules apply to two sequential
interactions, and to two parallel interactions.  The cat paper will be followed by two more papers that include
two additional rules that are specific to a continuous model of the brain \cite{RM2, RM3}.  The first of these
rules \{called rule (3a) in ref. 11\} is an extension of rule (3) that is specific to a continuous brain.  

The rules given in the present paper preserve the traditional epiphenomenal nature of consciousness, for
consciousness is not here given a causal role of any kind.  It simply appears when it is required to appear. 
However, the last of the continuous brain rules (called rule (5) in ref. 12) gives consciousness a causal role. 
It is structured in a way that allows von Neumann's \emph{psycho-physical parallelism} to naturally occur (ref.
8).  That is, the last rule enforces the \emph{parallel principle} that underlies and makes possible the evolution
of a psycho-physical parallelism in any conscious species \cite{RM4}.  As a result, consciousness has an effect
that is measurable in principle.  The author has already proposed an experiment that tests the parallel principle,
so it implicitly tests the structure defined by all of these rules \cite{RM5}.

\section*{Acknowledgements}

I wish to thank Fred Goldhaber, Erle Graf, Hal Medcalf, Luis Orozco, and       Greg Zelinsky for looking over this
paper and making useful comments and suggestions.

\end{document}